\documentclass[%
reprint,
superscriptaddress,
]{revtex4-1}
\usepackage{graphicx}
\usepackage{dcolumn}
\usepackage{bm}

\begin{document}

\title{Pulse Shape Discrimination of Fast Neutron Background using Convolutional Neural Network for NEOS II}
\author{Y.~Jeong}
\affiliation{Department of Physics, Chung-Ang University, Seoul, 06974, Korea}
\author{B.~Y.~Han}
\affiliation{Neutron Science Division, Korea Atomic Energy Research Institute, Deajeon, 34057, Korea}
\author{E.~J.~Jeon}
\affiliation{Center for Underground Physics, Institute for Basic Science (IBS), Daejeon, 34126, Korea}
\author{H.~S.~Jo}
\affiliation{Department of Physics, Kyungpook National University, Daegu, 41566, Korea}
\author{D.~K.~Kim}
\affiliation{Department of Physics, Kyungpook National University, Daegu, 41566, Korea}
\author{J.~Y.~Kim}
\affiliation{Department of Physics and Astronomy, Sejong University, Seoul, 05006, Korea}
\author{J.~G.~Kim}
\affiliation{Department of Physics, SungKyunKwan University, Suwon, 16419, Korea}
\author{Y.~D.~Kim}
\affiliation{Center for Underground Physics, Institute for Basic Science (IBS), Daejeon, 34126, Korea}
\affiliation{Department of Physics and Astronomy, Sejong University, Seoul, 05006, Korea}
\affiliation{IBS School, University of Science and Technology (UST), Daejeon, 34113, Korea}
\author{Y.~J.~Ko}
\affiliation{Center for Underground Physics, Institute for Basic Science (IBS), Daejeon, 34126, Korea}
\author{H.~M.~Lee}
\affiliation{Neutron Science Division, Korea Atomic Energy Research Institute, Deajeon, 34057, Korea}
\author{M.~H.~Lee}
\affiliation{Center for Underground Physics, Institute for Basic Science (IBS), Daejeon, 34126, Korea}
\affiliation{IBS School, University of Science and Technology (UST), Daejeon, 34113, Korea}
\author{J.~Lee}
\affiliation{Center for Underground Physics, Institute for Basic Science (IBS), Daejeon, 34126, Korea}
\author{C.~S.~Moon}
\affiliation{Department of Physics, Kyungpook National University, Daegu, 41566, Korea}
\author{Y.~M.~Oh}
\affiliation{Center for Underground Physics, Institute for Basic Science (IBS), Daejeon, 34126, Korea}
\author{H.~K.~Park}
\affiliation{Department of Accelerator Science, Korea University, Sejong, 30019, Korea}
\author{K.~S.~Park}
\affiliation{Center for Underground Physics, Institute for Basic Science (IBS), Daejeon, 34126, Korea}
\author{S.~H.~Seo}
\affiliation{Center for Underground Physics, Institute for Basic Science (IBS), Daejeon, 34126, Korea}
\author{K.~Siyeon}\email{siyeon@cau.ac.kr}
\affiliation{Department of Physics, Chung-Ang University, Seoul, 06974, Korea}
\author{G.~M.~Sun}
\affiliation{Neutron Science Division, Korea Atomic Energy Research Institute, Deajeon, 34057, Korea}
\author{Y.~S.~Yoon}
\affiliation{Center for Ionizing Radiation, Korea Research Institute of Standards and Science, Daejeon, 34113, Korea}
\author{I.~Yu}
\affiliation{Department of Physics, SungKyunKwan University, Suwon, 16419, Korea}

\collaboration{The NEOS II Collaboration}

\begin{abstract}
Pulse shape discrimination plays a key role in improving the signal-to-background ratio in NEOS analysis by removing fast neutrons. Identifying particles by looking at the tail of the waveform has been an effective and plausible approach for pulse shape discrimination, but has the limitation in sorting low energy particles. As a good alternative, the convolutional neural network can scan the entire waveform as they are to recognize the characteristics of the pulse and perform shape classification of NEOS data. This network provides a powerful identification tool for all energy ranges and helps to search unprecedented phenomena of low-energy, a few MeV or less, neutrinos.
\begin{description}
\item[Keywords]
reactor antineutrino, inverse beta decay, fast neutron, convolutional neural network, pulse shape discrimination
\end{description}
\end{abstract}

\maketitle

\section{\label{sec:level1}Introduction}
The existence of the fourth massive neutrino is a candidate solution to the lack of antineutrino flux measured in short distance from reactor cores \cite{Mention:2011rk}. Reactor neutrino experiments measuring the mixing angle $\theta_{13}$ commonly showed anomalous 5-MeV bumps in the spectral shape, also which are considered resolvable also with sterile neutrinos \cite{Adamson:2016jku}\cite{RENO:2018pwo}\cite{Adamson:2020jvo}. Neutrino Experiment for Oscillation at Short baseline(NEOS) is an electron antineutrino disappearance experiment to see oscillating aspect due to a 1-eV order neutrino mass, so-called light sterile neutrino.
The NEOS collected data from July 2015 to May 2016 including reactor-on 180 days and reactor-off 46 days, and reported the result with no strong evidence of a light sterile neutrino \cite{Ko:2016owz}\cite{Oh:neutrino2018}. After NEOS, more short-baseline experiments added exclusion areas in
$\Delta m_{14}^2$-$\sin^2{\theta_{14}}$ \cite{Ashenfelter:2018iov}\cite{Abreu:2017bpe}\cite{Alekseev:2018efk}\cite{Almazan:2018wln}.

NEOS II launched in September, 2018 with the same instrument setup as NEOS \cite{Ko:neutrino2020}\cite{Ko:2019cip}, and keeps taking data until November, 2020. This period covers about a 500-day cycle of fuel burn up, so full-time information available from a low-enriched uranium reactor allows to study changes in flux with fission fraction. Different composition of the isotopes during the cycle helps to extract the $\bar{\nu_e}$ flux of each single isotope, e.g.,$^{235}$U and $^{239}$Pu. The fission chain itself, which produces the reactor antineutrino flux, may be the cause of 5-MeV anomaly \cite{Dwyer:2014eka}. Or the nuclear properties may be a prerequisite for sterile neutrino to be an answer to the anomaly \cite{Adey:2019ywk}. NEOS II has a longer period of reactor shutdown for background measurements than NEOS. The rate+shape analysis will be added to the spectral shape analysis, and the results will be improved with the contribution of rich statistics and deep learning technology\cite{Kim Jinyu}.

Pulse-shape discrimination(PSD) is a step that can visually improve the signal-to-background ratio of detection.
NEOS focused on the attenuation part of the waveform that shows the shape of an event pulse moving over time. The charge fraction of pulse tail denoted by $F_\mathrm{tail}$ was used as a discriminating parameter to separate events by external fast neutrons from events by gammas and electrons \cite{Kim:2015pba}\cite{Ko:2016xuh}. On the other hand, convolutional neural network(CNN) is a deep learning method that has the advantage of recognizing and analyzing visual patterns \cite{Griffiths:2018zde}\cite{Holl:2019xtt}\cite{Abi:2020xvt}\cite{Alonso-Monsalve:2020nde}. Here we let the network take pulses of the entire shape, not just the tail or head, without pre-manipulating the shape.
It turns out that the neutron-gamma discriminant power of the network approach is enhanced so that more than 99.9\% of neutrons and gammas belong to neutron scores of 1 and 0, respectively.
Compared to $F_\mathrm{tail}$, CNN provides superior performance across all energy range. The inevitable blindness at low energy shown in the $F_\mathrm{tail}$ method is not a problem when the CNN drives PSD for the neutrons and gammas identification in NEOS detector.

The following section provides a brief overview of NEOS II, including detector specifications and IBD signal selection criteria. The next section describes the identification of particles in a test sample by two methods, $F_\mathrm{tail}$ and CNN. A reference sample is given for training, which is independent of the test sample. The final section discusses the robust discrimination obtained with CNN as a result and the possible extension of the task for future analysis.

\section{NEOS Detector and Data Acquisition}

\begin{figure}[t]
\centering
\includegraphics[width=0.45\textwidth]{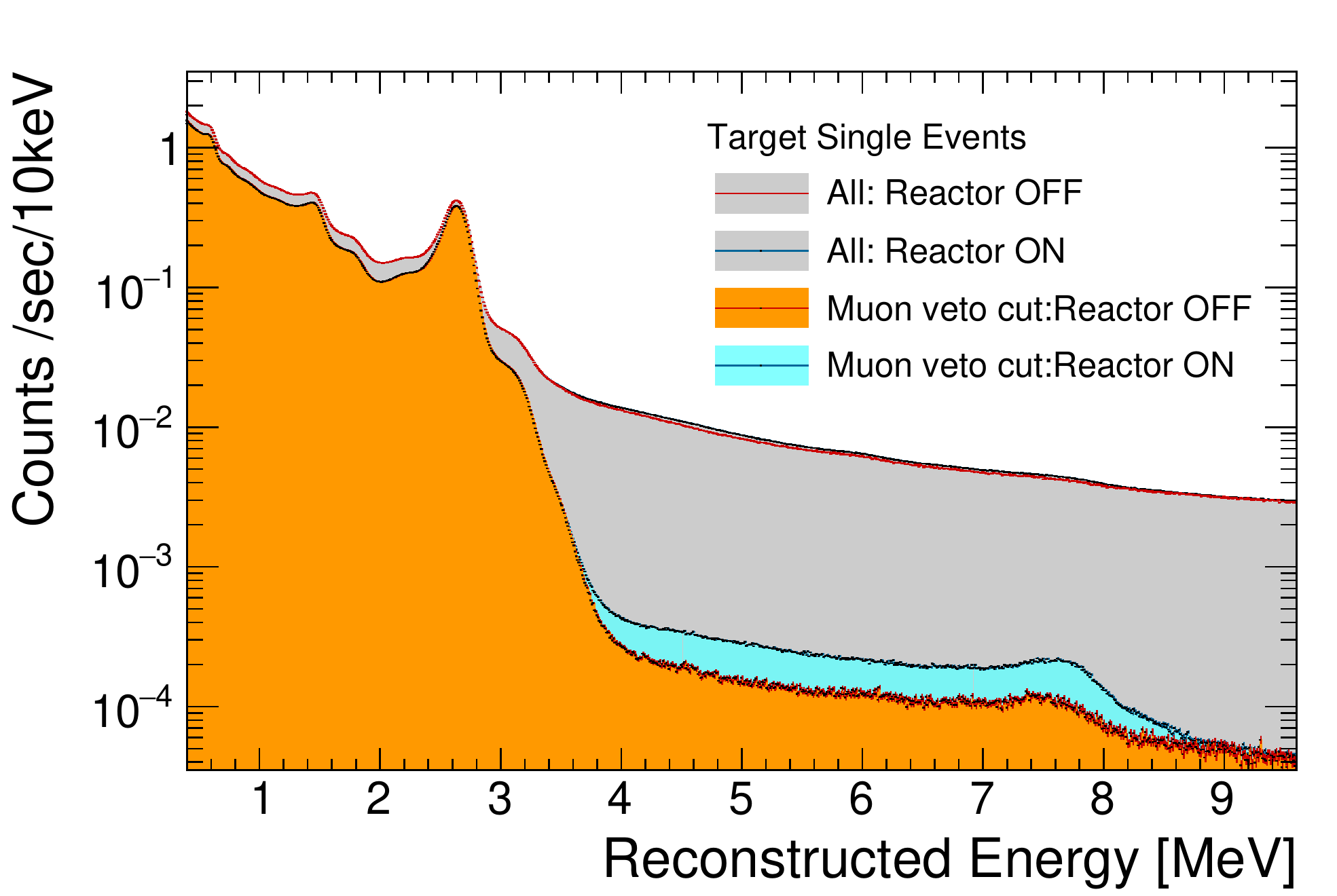}
\caption{Single-event spectrum of NEOS \cite{Oh:neutrino2018}. Without muon veto cut, turning the reactor on or off did not affect the event rate in NEOS detector. With muon veto cut, the difference between on and off(cyan area) between 3.5 MeV and 9 MeV shows the IBD signals of $\bar{\nu_e}$ originated from the reactor.} \label{fig:singles_neos1}
\end{figure}

NEOS is a 24-meter-baseline antineutrino oscillation experiment with a detector installed in reactor unit 5 of Hanbit Nuclear Power Complex in Yeonggwang, Korea. The overburden of the tendon gallery corresponds to a depth of 20 meters water equivalent \cite{Ko:2016owz}. A gadolinium(Gd)-doped(0.5\%) liquid scintillator(GdLS) 1008 liters uniformly filled the internal volume of the detector. A small amount of di-iso-propylnaphthalene(DIN)-based scintillation element(UG-F) was mixed with linear-alkyl benzene(LAB)-based GdLS to improve the light yield of the scintillator. The concentration, UG-F:LAB-GdLS=10:90, was determined to optimize the figure of merit in neutron-gamma discrimination \cite{Kim:2015pba}. The cylindrical LS container is 121 cm long and 104 cm in diameter. At each end of the vessel, nineteen 8-inch photomultiplier tubes(PMTs) are attached and are surrounded by a buffer of mineral oil. The additional shields consist of a 10-cm thick layer of borate polyethylene(B-PE) and lead(Pb), screening neutrons and gammas from external sources. Plastic scintillator boards as muon counters surround the outermost detector. NEOS II uses the detector of NEOS in the same environment.

The electron antineutrino is detected with inverse beta decay(IBD); $\bar{\nu}_e+p\rightarrow e^++n$. The prompt signal($S_1$) is gammas obtained from $e^-e^+$ annihilation, and the delayed signal($S_2$) is the collection of gamma cascade from the neutron capture by Gd. An IBD event candidate is selected by a coincident pair of signals, $S_1$ and $S_2$. If the prompt signal triggers data acquisition at $\mathrm{T_{prompt}}$, the delayed signal should be monitored in the 1 to 30 $\mu$s time window. The energy range of $S_1$ is 1 to 10 MeV, and its lower bound is limited to $2m_e$. $S_2$ is the gamma cascade of 8-MeV energy emitted from neutron-Gd capture. Fig.\ref{fig:singles_neos1} shows the record of all single events possible in NEOS detector \cite{Oh:neutrino2018}. With muon veto cut, IBD events with $S_2$ appear with a bump near 8 MeV and a reduction in energy due to an unavoidable gamma escape from the small-size detector. The selection criteria for $S_2$ has been extended to a range of 4 to 10 MeV. After all IBD selection criteria of energy and timing except PSD have been applied, most of the background is fast neutrons induced from cosmic muons, scattered by protons and then captured by Gd \cite{Ko:2016owz}.

Data is acquired by 500 mega-sampling-per-second flash analog-to-digital converter(FADC), and the signals from 38 8-inch PMTs are saved as waveforms in a 480-ns time window. Waveforms with well-defined pulse times are accumulated and form a synchronized waveform \cite{Ko:2019cip}. The events of recoil protons by elastic scattering of fast neutrons may mimic the IBD prompt signals. Within a period of hundreds nanoseconds, neutrons should be discriminated from photons. In Fig.\ref{fig:singles_neos1}, the difference part(cyan) between the reactor on/off represents mainly $S_2$ events for IBD pairs. Extracting the background neutrons from the orange area is what the PSD aims for in this work.

\section{Pulse Shape Discrimination}
\subsection{$F_\mathrm{tail}$ comparison}

The waveform is obtained from a pulse consisting of the rise, peak, and attenuation of the fluorescence generated from the excitation of electrons. The pulse shape depends on the type of particle that induces the excitation. For example, there are neutrons for proton recoil and photons for electron recoil. A waveform comparison technique for identifying scattering particles is pulse shape discrimination(PSD).
NEOS used a charge-to-charge fraction $F_\mathrm{tail}$ comparison as PSD and significantly removed the background fraction 73\% in the phase I \cite{Ko:2016owz}. The $F_\mathrm{tail}$ is defined as
\begin{eqnarray}
F_\mathrm{tail}\equiv\frac{Q_\mathrm{tail}}{Q_\mathrm{total}}
=\frac{\int_{T_c}^{T_\mathrm{max}} V(t)dt}{\int_{T_\mathrm{min}}^{T_\mathrm{max}} V(t)dt},
\end{eqnarray}
where $T_c$ is the start time of the tail part. $T_\mathrm{min}$ is the pulse threshold time and $T_\mathrm{max}$ is the signal gate duration.  The $F_\mathrm{tail}$ of the synchronized waveforms from 38 PMT has been evaluated with $T_\mathrm{max}-T_\mathrm{min}=340$ ns and $T_c-T_\mathrm{min}=100$ ns for NEOS II.

\begin{figure}[t]
\includegraphics[width=0.48\textwidth]{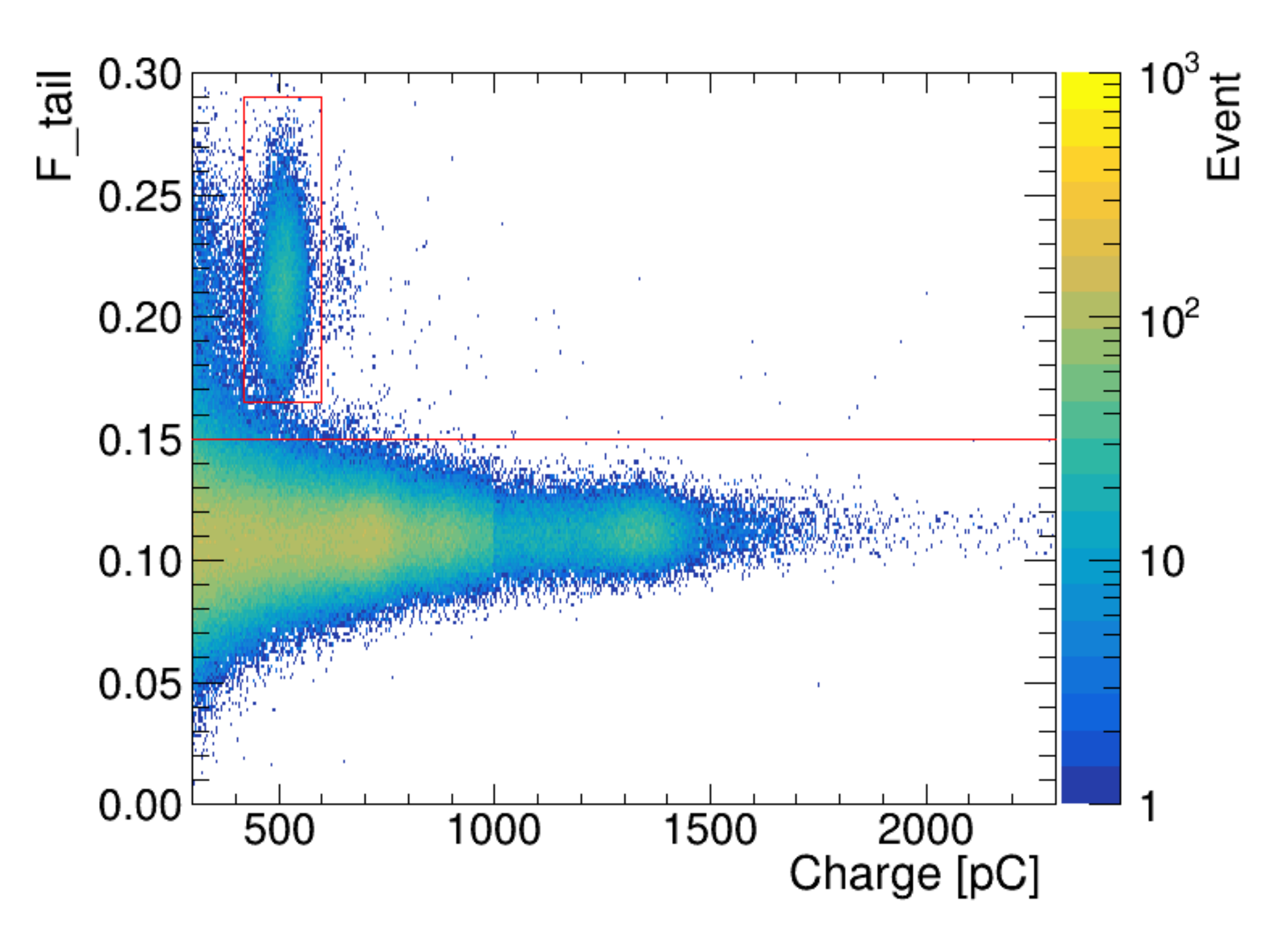}
\caption{\label{fig:fig2_AlphaSample} The $F_\mathrm{tail}$ of alpha and beta events in a reference sample. The sample contains 490,000 events extracted from NEOS-II physics run no. 270. The island around 500pC and of $F_\mathrm{tail}$ between 0.165 and 0.29 is the alpha events from $^{214}\mathrm{Po}$. The events selected by a cut $ F_\mathrm{tail}<0.15$ are of the beta decay from $^{214}\mathrm{Bi}$.}
\includegraphics[width=0.48\textwidth]{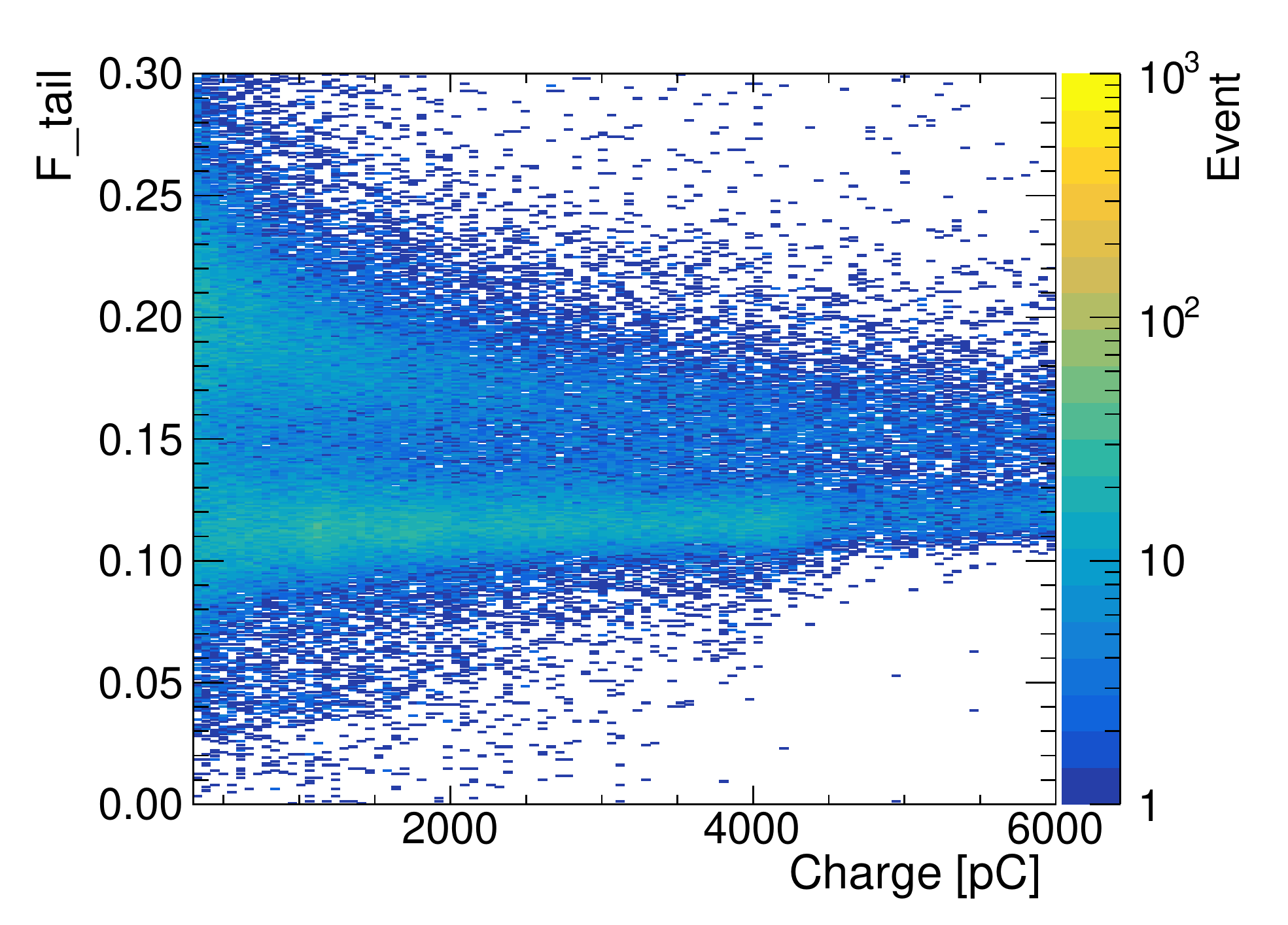}
\caption{\label{fig:fig4_CnnVsQtail}The $F_\mathrm{tail}$ of neutron and gamma events in a test sample. The sample contains 88,000 events extracted from NEOS-II physics run no. 270. The group of higher $F_\mathrm{tail}$ is neutrons, while the group of lower $F_\mathrm{tail}$ is gammas. It is a set of events collected in 150 $\mu$s after a muon-veto signal triggers, especially for the test of fast neutron and gamma discrimination. }
\end{figure}

For the PSD, both reference sample and test sample were collected from the NEOS-II physics run no.270. NEOS detector has coincident signal pairs combining the beta decays of $^{214}\mathrm{Bi}$ and the alpha decays of $^{214}\mathrm{Po}$
caused by $^{222}\mathrm{Rn}$ contamination of LS. It turns out that the alpha and beta events have the $F_\mathrm{tail}$ distributions well-separated as shown in Fig.\ref{fig:fig2_AlphaSample}, and about 490,000 events provide a powerful reference of waveforms for PSD. The events of an island in the box present the waveforms of alphas, while the events below $F_\mathrm{tail}=0.15$ present the waveforms of betas.

\begin{figure}[t]
\centering
\includegraphics[width=0.45\textwidth]{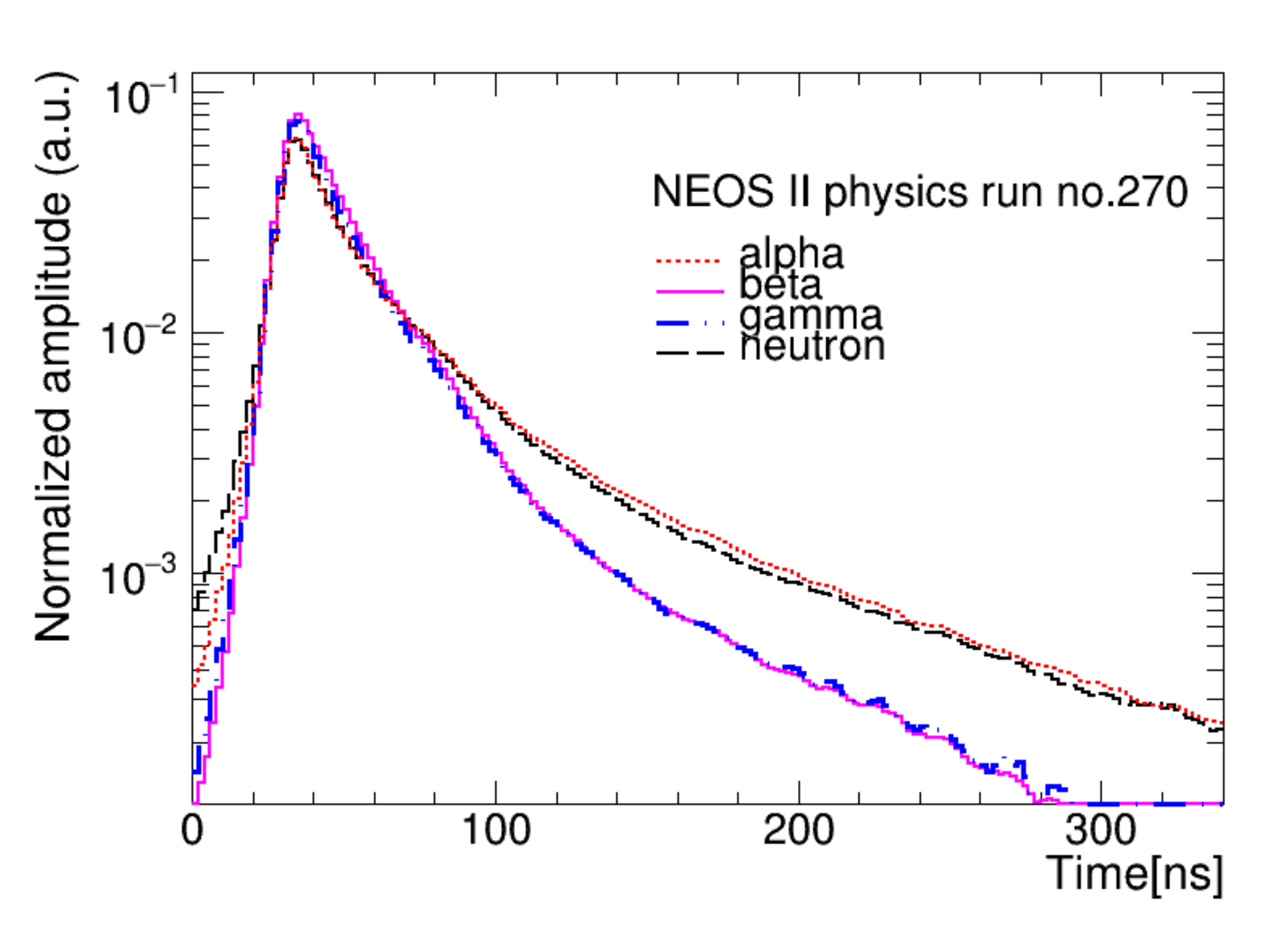}
\caption{\label{fig:fig1_SyncWave} Comparison of the pulse shapes. For each curve, a number of 38-PMT-synchronized waveforms are accumulated and the area is normalized to one. Selected events for the accumulation are 16,000 alphas and 13,000 betas from Fig.\ref{fig:fig2_AlphaSample}, and 1,900 gammas and 4,200 neutrons from Fig.\ref{fig:fig4_CnnVsQtail}. The shapes of neutrons(black dashed line) and gammas(blue dashed-dotted line) are regarded as the same as the shapes of alphas(red dotted line) and betas(pink solid line), respectively.}
\end{figure}

The figure \ref{fig:fig4_CnnVsQtail} is another $F_\mathrm{tail}$ distribution for PSD. About 88,000 events detected within 150 $\mu$s after a muon-veto event beeps were selected as test sample of fast neutrons and gammas. Referring to the $F_\mathrm{tail}$ of alphas and betas in Fig.\ref{fig:fig2_AlphaSample}, particles in waveform of alphas are counted as neutrons, while particles in waveform of betas are counted as gammas. Then, the events appear as divided into two groups in terms of $F_\mathrm{tail}$. The $F_\mathrm{tail}$ is still blind to identifying neutrons and gammas in the region below 2000 pC near $F_\mathrm{tail}=0.15$.

The figure \ref{fig:fig1_SyncWave} explains why the neutrons and gammas in the test sample can take the $F_\mathrm{tail}$ of alphas and betas in the reference sample. It displays the waveform accumulations of four types of particles, alpha, beta, gamma and neutron by red(dotted), red(solid), blue(dashed-dotted) and black (dashed) curves, respectively. 16,000 events of alphas and 13,000 events of betas in Fig.\ref{fig:fig2_AlphaSample} and 4,200 events of neutrons and 1,900 events of gammas in Fig.\ref{fig:fig4_CnnVsQtail} were selected for the accumulation. The curve of each particle is normalized to one equal area. When comparing the shape of the pulse, it was found that neutrons and alphas are indistinguishable and so are gammas and betas, justifying that the waveforms of alpha and beta can replace the waveforms of neutron and gamma, respectively.

\subsection{Convolutional neural network}

The pulse shape discrimination in the framework of CNN performs two independent processes in stage: one is training CNN to learn waveforms from the reference sample of 490,000 events in Fig.\ref{fig:fig2_AlphaSample} and the other is identifying neutrons and gammas from the features of the 88,000 events of the test sample. The neutron and gamma candidates are the same events used in Fig.\ref{fig:fig4_CnnVsQtail}, but CNN reads the entire waveform as it is without being dissected with the head or tail.
The energy distribution of alpha and beta events obtained from Bi and Po sources is less broad and limited to the lower region than the energy region 460 pC to 6000 pC of the neutron and gamma events. The energy of alpha is 400-600 pC, and the energy of beta does not exceed 2000 pC at most. That is, the training reference is the events of lower energy than the events of IBD channels.

\begin{figure}[t]
\includegraphics[width=0.30\textwidth]{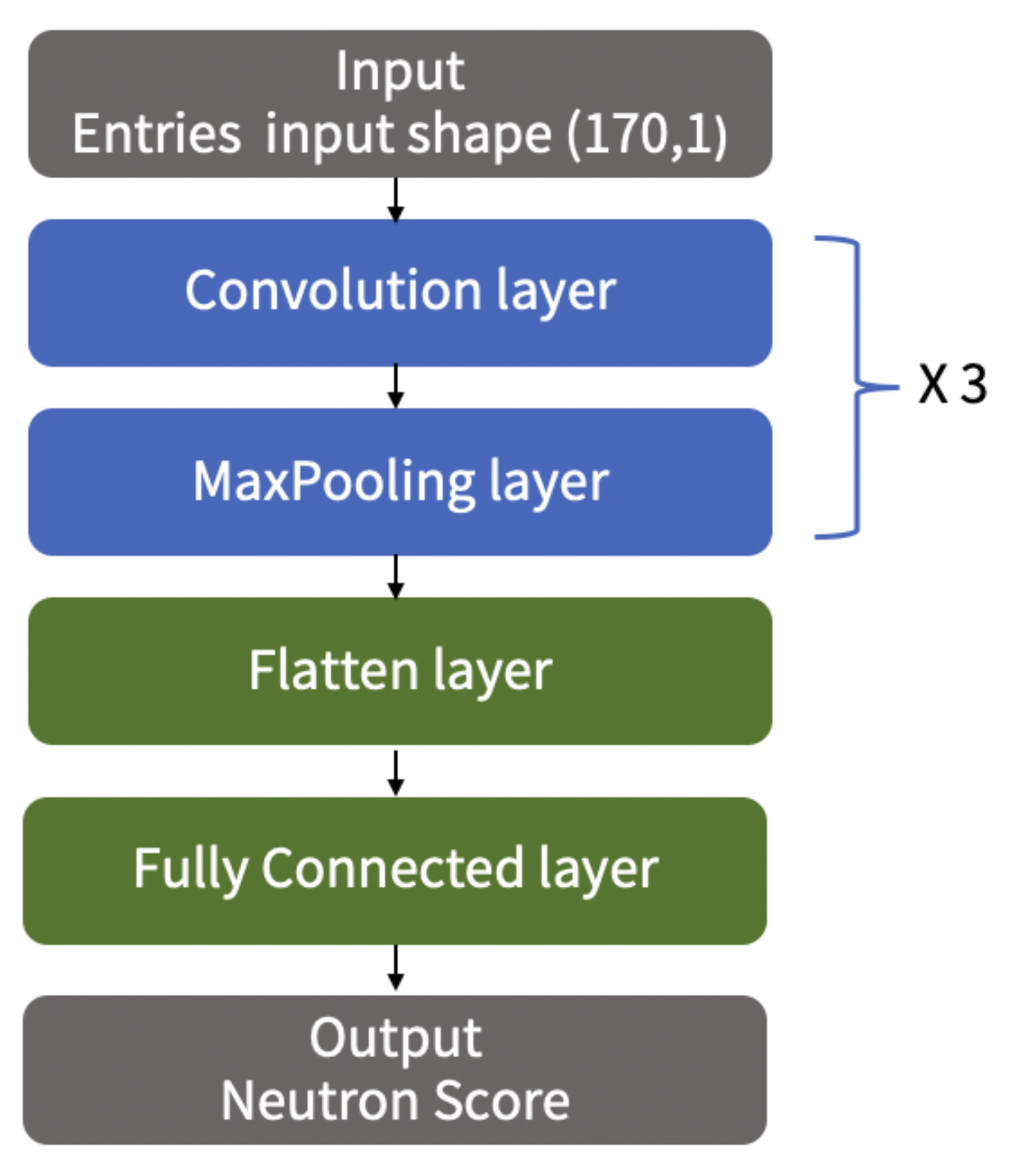}
\caption{\label{fig:fig3_NetworkFlow} CNN architecture of particle discrimination. The input is provided by 170 1D data representing the synchronized waveform of an event and by a batch size representing the number of events. After iterations of the convolution and maxpooling routines to extract features of the data, the shape of the data is flattened and entered into a fully-connected neural network. The activation function sigmoid is adopted for the binary classifier to identify neutron or gamma in terms of the neutron score.}
\end{figure}

\begin{figure}[t]
\includegraphics[width=0.48\textwidth]{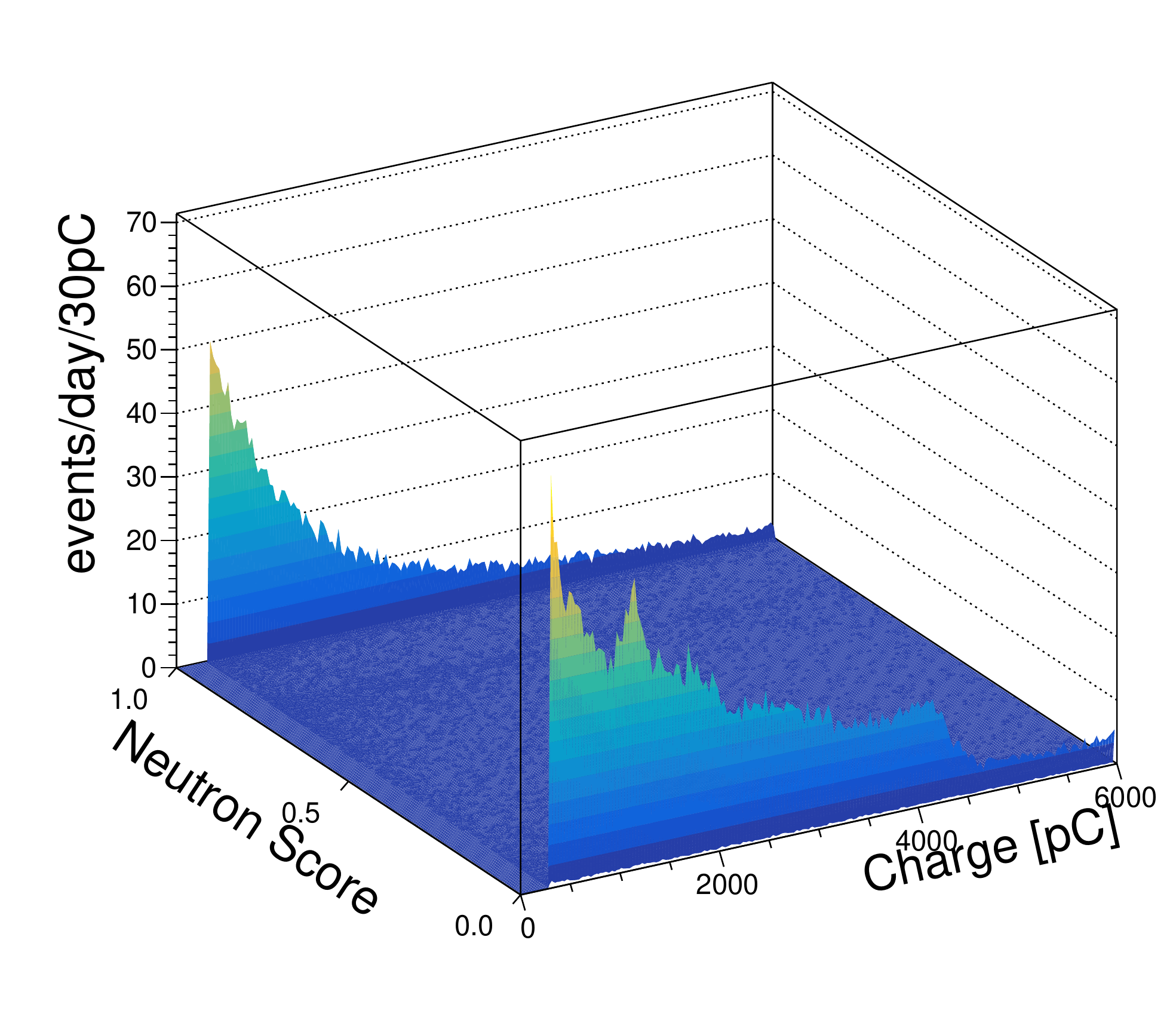}
\caption{\label{fig:fig5_Nscore} Discrimination of neutron and gamma events using CNN. The 3-dimensional display of event distribution for the neutron score and the charge illustrate completely separated spectrum, the neutron at score 1 and the gamma at score 0.}
\includegraphics[width=0.45\textwidth]{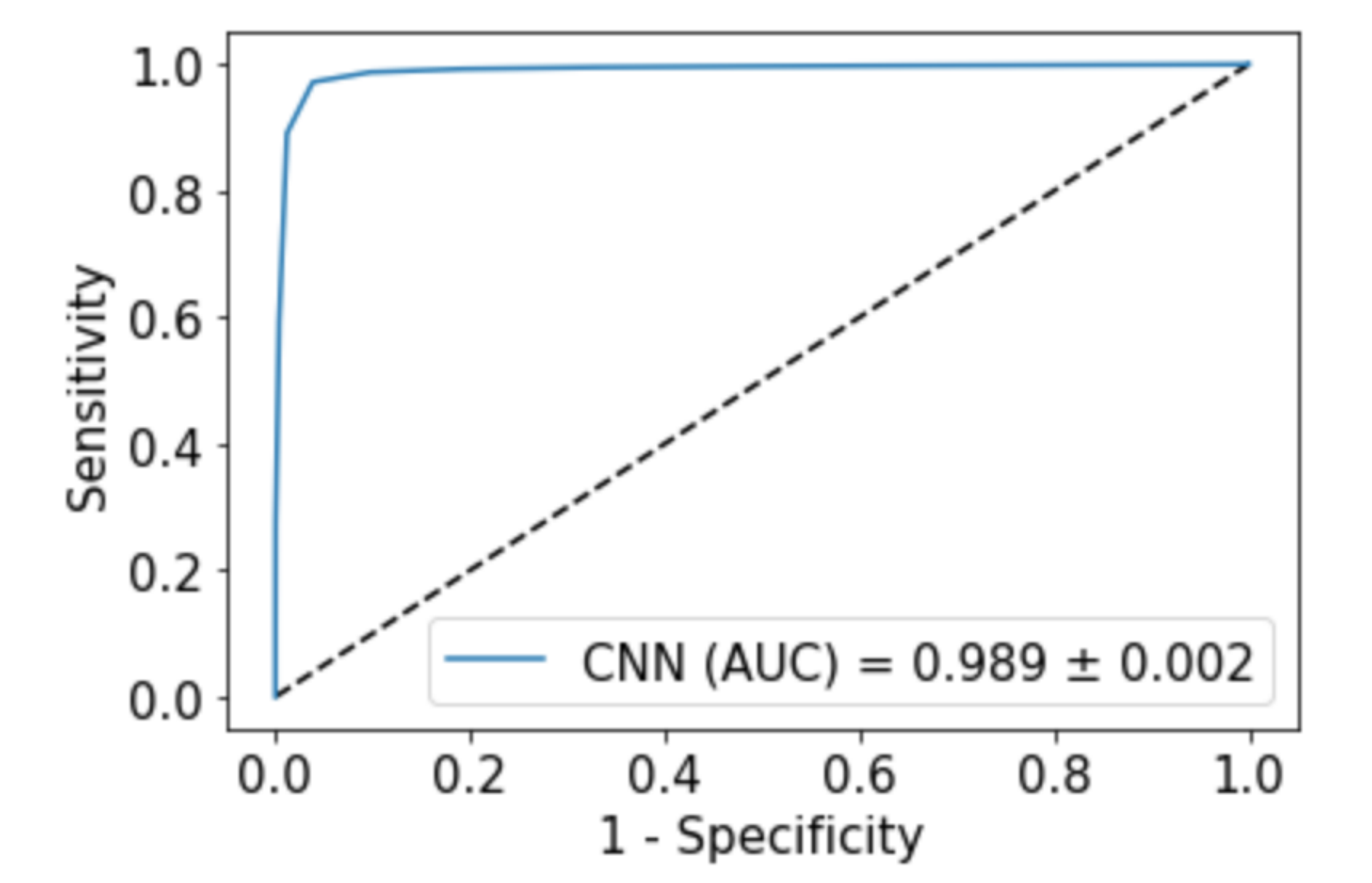}
\caption{\label{fig:fig6_RocAuc} ROC curve of the CNN for the neutron-gamma PSD. The AUC 0.989 indicates that CNN has reached a high level of discrimination accuracy. }
\end{figure}

The structure of CNN is composed of the feature extraction from the input data and the classification of data via neural network, as shown in Fig.\ref{fig:fig3_NetworkFlow}. Convolution and max-pooling layers are alternately repeated to extract features. A one-dimensional array of 170 numbers represents a synchronized 340-ns waveform of an event and becomes an input to 64 three-kernel filters of convolution layer, and significant features are extracted in the max-pooling stage to improve the operation. After the sequence of convolution and pooling was repeated three times, the shape of output is modified in flatten layer to be the shape of input adequate to fully-connected neural network. The fully-connected layer consists of hidden layers and activation functions, which finally assign a particle the score as classification.

Activation functions working between hidden layers in the fully-connected neural network determine the output score. The activation `sigmoid' is the one adopted for binary classifier, e.g., true and false, 1 and 0, or neutron and gamma. While the training process with the reference sample runs, weight and bias of sigmoid are optimized so that neutron-like and gamma-like inputs result in the scores 1 and 0, respectively. After the CNN for PSD runs 490,000 events, it is trained such that the output of activation sigmoid approaches to 1 or 0, and the weight and the bias is finally fixed. Then, the CNN runs with 88,000 events of the test sample. The function evaluated with a test event outputs a CNN score scaled between 0 and 1, so that it can be interpreted as the probability of a neutron event. Since the score 1 represents a 100\% probability of identifying a neutron, we named the value `neutron score'. A neutron score 0 means an event with 100\% probability of gamma.

The figure \ref{fig:fig5_Nscore} and the figure \ref{fig:fig6_RocAuc} illustrate the result of CNN to discriminate neutrons from the test sample of NEOS II and its Receiver Operating Characteristic(ROC) curve, respectively. More than 99.9\% of the neutrons and gamma events are identified by neutron scores 1 and 0, respectively. The discriminating power is high over all energy range, and its accuracy is described in terms of the sensitivity of score-1 neutrons and the specificity of score-0 gammas, which measure the proportion of the identified particles predicted from the model to the total identified particles either predicted or unpredicted from the model. The area under ROC curve(AUC) 0.989 indicates almost maximal sensitivity and specificity.

\section{Result and Discussion}

The separation between neutrons and gammas is significantly improved using CNN. More than 99.9\% of the events belong to the neutron score of 1 or 0. Therefore, both discriminating power and its accuracy have shown better results compared to the previous single parameter method.

In the Fig.\ref{fig:fig5_Nscore}, the distribution of events at neutron score zero corresponds to the spectrum of gamma events. As explained for the NEOS detector in Fig.\ref{fig:singles_neos1}, the 8-MeV bump of the single-event spectrum is IBD $S_2$(8 MeV) characterizing the gamma spectrum. In fact, the position of a particular peak at 4000 pC in the spectrum with neutron score 0 is the same as the single event spectrum in Fig.\ref{fig:singles_neos1}, considering the charge-to-energy conversion 500 pC per MeV. There is no peculiar point about the spectrum for neutron score 1.

Additionally, the particle discrimination in CNN provides an access to the low energy gamma events, while the previous $F_\mathrm{tail}$s of neutrons and gammas overlap at low energies, even if the reference sample provides a clean boundary between them. Therefore, the CNN approach could bring effectiveness in understanding the low-energy aspects of neutrons and gammas so that the signal-to-background ratio can be improved for the NEOS II analysis.

In this work, a test sample for $F_\mathrm{tail}$ and CNN was selected with events occurring near muon detector signal. The analysis focused on the case where the portion of fast neutrons was relatively higher than the entire event content to investigate discrimination. The particle identification using CNN will be applied to a single-event spectrum for the upcoming NEOS II neutrino-oscillation analyses.

\begin{acknowledgments}
Y. Jeong and K. Siyeon thank C. Ha for helpful advice and discussion. This research was supported by the National Research Foundation Grant of Korea (NRF-2017R1A2B4004308), IBS-R016-D1, and the Chung-Ang University Graduate Research Scholarship in 2019.
\end{acknowledgments}

\end{document}